\documentclass[a4paper,11pt]{article}
\usepackage{caption}
\usepackage{epsfig, graphicx, subfloat, float}
\usepackage{latexsym}
\usepackage{subcaption}
\usepackage{cite}
\usepackage{amssymb}
\usepackage{latexsym}
\usepackage{mathrsfs}
\usepackage{amsmath}
\usepackage{epstopdf}
\usepackage{color}
\usepackage[usenames]{xcolor}

\author{H. Mohseni Sadjadi \footnote{mohsenisad@ut.ac.ir}
\\ {\small Department of Physics, University of Tehran,}
\\ {\small P. O. B. 14395-547, Tehran 14399-55961, Iran}}
\title {Scalar-Gauss-Bonnet model, the coincidence problem and the gravitational wave speed}
\begin{document}
\maketitle
\begin{abstract}
We introduce a dynamical dark energy model wherein quintessence interacts with both the Gauss-Bonnet invariant and dark matter. Initially, the Gauss-Bonnet invariant stabilizes the quintessence at a fixed point, resulting in a negligible density of dark energy. Subsequently, the conformal coupling to dark matter triggers the evolution of dark energy. This model proposes an explanation for the initial absence of dark energy in radiation era and its later emergence during the matter-dominated era, achieving a magnitude comparable to dark matter in the present epoch. In this scenario, the Gauss-Bonnet term does not directly influence late-time cosmic evolution. Our model aligns with the assumption that the speed of gravitational wave is infinitesimally close to the speed of light.

\end{abstract}

\section{Introduction}

For over two decades, it has been established that the late-time expansion acceleration of the Universe is positive \cite{acc1,acc2}. Various models have been proposed to elucidate this phenomenon. One of the earliest and simplest models is the $\Lambda CDM$ model. Despite its alignment with many observational data, this model encounters fundamental issues such as fine-tuning \cite{ft}, and the coincidence problem\cite{cp}. Consequently, extensive research has been devoted to models involving exotic fields like dark energy or modifications and extensions of the standard model of gravity\cite{de1,de2,de3,de4,de5,de6,de7,de8,de9,de10}. In this context, many articles have employed the Gauss-Bonnet (GB) term to investigate the Universe's evolution \cite{GB,GB1,GB2,GB3,GB4}. This term, a total derivative in four dimensions, does not independently alter the Friedmann equations and thus does not directly influence the evolution of the Universe. Consequently modified Gauss-Bonnet \cite{mgb1,mgb2,mgb3} or generalized models incorporating a scalar field coupled to the GB term (SGB model) have been utilized to study cosmic acceleration \cite{SGB1,SGB2,SGB3,SGB4,SGB5,SGB6,SGB7,SGB8,SGB9}. Additionally, this model has garnered significant attention in the physics of black holes due to its provision of conditions for scalarization \cite{sc1,sc2}. Scalarization in cosmology was also studied in \cite{sc3,sc4}.

The detection of gravitational wave resulting from the merger of two neutron stars (GW170817 event) and their corresponding electromagnetic counterpart \cite{GW1} may establish observational conditions on the speed of gravitational wave at low redshifts \cite{GW2}. These constraints significantly limit the parameter space and dynamics of fields within modified gravity models like the Scalar-Gauss Bonnet (SGB) model, which predict gravitational wave speed differing from that of light \cite{GWL1,GWL2,GWL4,GWL5,GWL6,GWL7,GWL8,GWL9,GWL10}. Adhering to these speed constraints within the SGB model might exclude the Gauss-Bonnet term from contributing to late-time expansion as claimed in \cite{GWL3,Shin}. While the study \cite{GW2} and subsequent articles have suggested that constraints within Horndeski's theories rule out the direct contribution of the Scalar-Gauss Bonnet model to late-time acceleration, the issue is somewhat nuanced. An argument presented in \cite{Cl} and elucidated using a simplified model possessing a known partial UV completion \cite{Cl1} suggests that at energy scales observed by  LIGO, close to the cutoff linked with a dark energy model, gravitational wave might not necessarily propagate at speeds different from that of light. Within the Horndeski framework,  \cite{Cl} demonstrates that inherent operators existing at the cutoff scale could potentially restore gravitational wave speed propagation to that of light at LIGO scales. In our study, we assume that the speed of gravitational wave is very close to the speed of light, especially at low redshifts. This assumption is tuned to align with the narrow domain specified in \cite{GW2}.

Since dark energy dilutes more slowly than matter and radiation, we expected that dark energy would fully dominate the universe, while dark energy and matter have the same order of magnitude today. By arguing that dynamical dark energy was vanishingly insignificant initially, but has since grown and become significant in the matter-dominated era, one can alleviate this coincidence problem. In addition, observations, particularly in structure formation, suggest that the quantity of dark energy in the early universe should have been relatively small or even negligible (see \cite{EDE1,EDE2,EDE3,EDE4} and related references). Therefore, it becomes intriguing to develop a formal framework that explains why dynamical dark energy remained negligible during the early eras.

In this paper, our aim is to propose a model that elucidates the absence of dark energy in the early universe, aligning with the findings by \cite{GWL3,Shin}. We investigate cosmic acceleration within the framework of the SGB (or quintessence-GB) model, incorporating a dark matter component conformally coupled to the quintessence, which we take as a mass-varying dark matter.
In our study, the GB term doesn't directly contribute to the late-time expansion. Instead, it fulfills a crucial role: initially maintaining the quintessence at a stable fixed point with zero dark energy density. During this phase, the Universe's evolution obeys the standard Friedmann equations, encompassing only ordinary components . As the Universe expands and radiation and matter dilute, the initial point becomes unstable due to the coupling between quintessence and dark matter. This instability triggers the evolution of quintessence, leading to the emergence of dark energy during the matter-dominated era. This mechanism bears resemblance to cosmological scalarization, wherein the scalar field becomes tachyonic and acquires a nontrivial solution \cite{sc3}. The scheme of the paper is as follows:

In the next section, we introduce the model and show how by appropriable selection of parameters the GB term keeps the Universe in a state with negligible dark energy density initially. The role of the conformal coupling and the quintessence potential to onset of dark energy is explained. We illustrate our results through a numerical example, which allows us to select a tiny GB coupling and to have a slowly varying quintessence at the late time, playing the role of a cosmological constant. We conclude our results in the third section.

Throughout the paper, we use natural units $\hbar=c=1$.

\section{Scalar-Gauss-Bonnet model and onset of dark energy}
We consider the Scalar-Gauss-Bonnet (SGB) action
\begin{equation}\label{1}
S=\int d^4x \sqrt{-g}\left(\frac{M_P^2 R}{2}-\frac{1}{2}g^{\mu \nu}\partial_\mu \phi \partial_\nu \phi-V-\frac{1}{2}f\mathcal{G}\right)+S_m,
\end{equation}
where $M_P$ is the reduced Planck mass, $R$ is the Ricci scalar, and $\mathcal{G}=R_{\mu \nu \rho \sigma}R^{\mu \nu \rho \sigma}-4R_{\mu \nu}R^{\mu \nu}+R^2$ is the Gauss-Bonnet (GB) invariant coupled to the quintessence through the real function $f=f(\phi)$. $\phi$ is a real scalar field with potential $V=V(\phi)$. The first part of the action is a special case of the Horndeski action \cite{hord}
\begin{eqnarray}\label{2}
&&S=\int d^4x \sqrt{-g} \Big[G_2-G_3\Box \phi(x)+G_4R+G_{4,X}\left((\Box \phi)^2-(\nabla_\mu \nabla_\nu \phi)(\nabla^\mu \nabla^\nu \phi)\right)\nonumber \\
&&-\frac{1}{6}G_{5,X}\left((\Box \phi)^3-3(\Box \phi)(\nabla_\mu \nabla_\nu \phi)(\nabla^\mu \nabla^\nu \phi)+2(\nabla^\mu \nabla_\alpha \phi)(\nabla^\alpha \nabla_\beta \phi)(\nabla^\beta \nabla_\mu \phi)\right) \nonumber \\
&& +G_5G_{\mu \nu}\nabla^\mu \nabla^\nu \phi \Big]+S_m,
\end{eqnarray}
with \cite{cob}
\begin{eqnarray}\label{3}
&&G_2=X-V-4f_{,\phi\phi\phi\phi}X^2(3-lnX)\nonumber \\
&&G_3=-2f_{,\phi\phi\phi}X(7-3lnX)\nonumber\\
&&G_4=\frac{M_P^2}{2}-2f_{,\phi\phi}X(2-lnX)\nonumber \\
&&G_5=2f_{,\phi} lnX,
\end{eqnarray}
where $X=-\frac{1}{2}g^{\mu \nu}\nabla_\mu \phi \nabla_\nu \phi$, $G_{\mu \nu}$ is the Einstein tensor, and  $f_{,\phi}=\frac{df}{d\phi}$.

$S_m$ contains matter ingredients like radiation, baryonic matter, and cold dark matter. We assume a dark matter component, indicated by a $\sigma$ index, interacts with the quintessence. This interaction is achieved by considering this component as a mass-varying dark matter whose mass is a function of $\phi$: $m_{\sigma}(\phi)$, such that \cite{mv1,mv2}
\begin{equation}\label{4}
\nabla_\mu {T^{(\sigma)}}^{\mu}_{\nu}=\frac{m_{\sigma ,\phi}}{m_\sigma} {T^{(\sigma)}}^\mu_{\mu} \phi_{,\nu},
\end{equation}
where ${T^{(\sigma)}}_{\mu \nu}$ is the stress tensor of this dark matter.

We consider the spatially flat (Friedmann-Lema\^{\i}tre- Robertson-Walker) FLRW space-time with the metric
\begin{equation}\label{5}
ds^2=-dt^2+a^2(t)(dx^2+dy^2+dz^2).
\end{equation}
In this space-time
\begin{equation}\label{6}
\mathcal{G}=24H^2(\dot{H}+H^2).
\end{equation}
A dot means a derivative with respect to cosmic time. We make the assumption that the background quintessence field is a homogenous, time-dependent scalar field: $\phi=\phi(t)$. By varying (\ref{1}) with respect to the metric, we obtain modified Friedmann equations
\begin{eqnarray}\label{7}
&&3M_P^2H^2=\frac{1}{2}\dot{\phi}^2+V+12H^3f_{,\phi}\dot{\phi}+\rho \nonumber \\
&&2(M_P^2-4H\dot{f})\dot{H}=-\dot{\phi}^2+4H^2(\ddot{f}-H\dot{f})-P-\rho.
\end{eqnarray}
$\rho=\Sigma_i \rho_i$ is the total energy density of barotropic perfect fluids, and $P=\Sigma_i P_i$ is their total pressure.
The continuity equations for interacting sectors are:
\begin{equation}\label{8}
\dot{\rho}_\sigma+3H\rho_\sigma=B_{,\phi}\dot{\phi}\rho_\sigma,
\end{equation}
and
\begin{equation}\label{9}
\ddot{\phi}+3H\dot{\phi}+V_{,\phi}+12H^2(H^2+\dot{H})f_{,\phi}=-B_{,\phi}\rho_\sigma,
\end{equation}
obtained by varying the action with respect to $\phi$. $B$ is defined as $B:=\ln m_\sigma$.
Note that similarly, we could obtain the above evolution equations by taking the formalism used in the screening model $S_m=\sum_i S_m(\tilde{g}^i_{\mu \nu}\Psi_i)$ where the metric in $S_m$ is conformally coupled to the scalar field such that for
each matter species $\psi_i$, the conformal coupling is $\tilde{g}^i_{\mu \nu}={A_i}^2(\phi)g_{\mu \nu}$ \cite{bean,sadj1,sadj2}. If one takes $A_i=1$, except for $\sigma$ dark component where $A_\sigma=A(\phi)$, then \cite{sym1,sadj1,sadj2}
\begin{eqnarray}\label{dm}
&&\dot{\rho}_\sigma+3H\rho_\sigma=\frac{A_\phi}{A}\dot{\phi}\rho_\sigma, \nonumber\\
&&\ddot{\phi}+3H\dot{\phi}+V_{,\phi}+12H^2(H^2+\dot{H})f_{,\phi}=-\frac{A_\phi}{A}\rho_\sigma,
\end{eqnarray}
which is the same as (\ref{8}) and (\ref{9}), provided that $B_{,\phi}=\frac{A_\phi}{A}$, or $A=\frac{m_\sigma}{m}$, where $m$ is a mass scale.

Radiation, baryonic, and non-coupled dark matter satisfy
\begin{eqnarray}\label{10}
&&\dot{\rho}_r+4H\rho_r=0\nonumber\\
&&\dot{\rho}_m+3H\rho_m=0.
\end{eqnarray}
By $\rho_m$, we denote the sum of baryonic and non-coupled dark matter energy densities. We could also define an effective dark energy density, and pressure by
\begin{equation}\label{ref1}
\rho_d=\frac{1}{2}\dot{\phi}^2+V+12H^3\dot{f},
\end{equation}
and
\begin{equation}\label{ref2}
P_d=\frac{1}{2}\dot{\phi}^2-V-8H^3\dot{f}-4H^2\ddot{f}-8H\dot{H}\dot{f},
\end{equation}
respectively, satisfying the continuity equation
\begin{equation}\label{ref3}
\dot{\rho_d}+3H(P_d+\rho_d)=-B_{,\phi}\dot{\phi}\rho_\sigma.
\end{equation}
In this way, the modified Friedmann equations reduce to
\begin{eqnarray}\label{ref4}
&&3M_P^2H^2=\rho_d+\rho_r+\rho_m+\rho_\sigma \nonumber \\
&&2M_P^2\dot{H}=-\rho_d-P_d-\rho_m-\rho_\sigma-\frac{4}{3}\rho_r. \nonumber
\end{eqnarray}

The speed of gravitational wave might be obtained by studying the tensor perturbation \cite{pert,pert1,pert2}. Perturbing the metric as
\begin{equation}\label{11}
ds^2=-dt^2+a^2(t)(\delta_{ij}+h_{ij})dx^i dx^j,
\end{equation}
where $h_{ij}$ is the traceless ($h^i_i=0$), and divergence-free ($\partial^i h_{ij}=0$) tensor perturbation.
Assuming that the gravitational wave propagates in
the $+z$ direction, and defining  $h_1:=h_{11}=-h_{22}$, and $h_2:=h_{21}=h_{12}$, by expanding the action up to the second order of $h_i$, one obtains \cite{pert1}
\begin{equation}\label{12}
S_T^{(2)}=\frac{1}{4}\int d^4x a^3\sum_{i=1,2}q_T\left(\dot{h_i}^2-\frac{c_T^2}{a^2}(\partial h_i)^2\right),
\end{equation}
where
\begin{equation}\label{13}
q_T=2G_4-2\dot{\phi}^2G_{4,X}+\dot{\phi}^2G_{5,\phi}-H\dot{\phi}^3G_{5,X},
\end{equation}
with $X=\frac{1}{2}\dot{\phi}^2$. So, the gravitational wave speed is obtained as \cite{pert1,pert2}
\begin{equation}\label{15}
c_{T}^2=\frac{2G_4-\dot{\phi}^2G_{5,\phi}-\dot{\phi}^2G_{5,X}\ddot{\phi}}{q_T}
\end{equation}
In our model (1), after some computations, $q_T$ reduces to
\begin{equation}\label{14}
q_T=-4H\dot{\phi}f_{,\phi}+M_P^2.
\end{equation}
For (\ref{15}) we obtain
\begin{equation}\label{16}
c_{T}^2=\frac{4f_{,\phi\phi}\dot{\phi}^2+4f_{,\phi}\ddot{\phi}-M_P^2}{4Hf_{,\phi}\dot{\phi}-M_P^2}=
\frac{4\ddot{f}-M_P^2}{4H\dot{f}-M_P^2},
\end{equation}
where to derive the last equality we have used $\ddot{f}=f_{,\phi}\ddot{\phi}+f_{,\phi \phi}\dot{\phi}^2$. We assume that in our model, the speed of gravitational wave, $c_T$, is very close to the light speed, such that it falls into the range reported in
\cite{GW2} for late times ($z<0.009$):
\begin{equation}\label{17}
-3\times 10^{-15}\leq \frac{c_{T}}{c}-1 \leq 7\times 10^{-16},
 \end{equation}
 where $c$ is the light speed (to recover the speed unit, we must replace $c_T$ by $\frac{c_T}{c}$ in(\ref{16})).
 Using $\left(\frac{c_{T}}{c}\right)^2-1\simeq 2(\frac{c_{T}}{c}-1)$, which holds for $\frac{c_{T}}{c}\simeq1$, we obtain $|\left(\frac{c_T}{c}\right)^2-1|\lesssim 10^{-15}$ \cite{Shin}.
 This bound implies generally  $\ddot{f}\ll M_P^2$ and $4H\dot{f}\ll M_P^2$. Therefore, under the assumption (\ref{17}), we infer from (\ref{7}) that the direct contribution of the GB term to late-time acceleration is negligible compared to other terms. Despite this, as mentioned in the introduction, the GB term may have played an important role in earlier times by providing the conditions for the existence of an initial stable fixed point solution sans dark energy. However, owing to the conformal coupling, the system loses its stability at this point, leading to a dynamical quintessence. Consequently, dark energy becomes notably significant during the matter dominated era. This scenario offers an explanation for the absence of dark energy in the early universe and its ascendancy in the late-time. To model this formalism, we proceed as follows:

We take $f$, $V$, and $B$ as even functions of $\phi$
\begin{equation}\label{ref5}
f=f(\phi^2)\,\,\,  V=V(\phi^2),\,\,\ B=B(\phi^2).
\end{equation}
By this selection, the action (\ref{1}) gains $Z_2$ symmetry, i.e. is invariant under $\phi\leftrightarrow -\phi$. We rewrite (\ref{9}) as
 \begin{equation}\label{ref6}
 \ddot{\phi}+3H\dot{\phi}+V^{ef.}_{,\phi}=0,
 \end{equation}
 where the effective potential is given by \cite{Gom}
 \begin{equation}\label{ref7}
V^{ef.}_{,\phi}=V_{,\phi}+12H^2(H^2+\dot{H})f_{,\phi}+B_{,\phi}\rho_\sigma.
\end{equation}
As $f$, $V$, and $B$ are even functions, we have $V_{,\phi} (0)=f_{,\phi}(0)=B_{,\phi}(0)=0$, therefore $V^{ef.}_{,\phi}(0)=0$. We also choose a positive potential such that
$V(0)=0$.  Therefore $\phi=0$ is a {\it{trivial}} solution to the equation (\ref{ref6}), corresponding to zero dark energy density $\rho_d=0$, respecting the $Z_2$ symmetry. The Friedmann equations for this trivial solution reduce to the standard ones
\begin{eqnarray}\label{ref8}
H^2=\frac{1}{3M_P^2}\left(\rho_m+\rho_r+\rho_\sigma\right)\nonumber \\
\dot{H}=-\frac{1}{2M_P^2}\left(\rho_m+\rho_\sigma+\frac{4}{3}\rho_r\right).
\end{eqnarray}
 So from (\ref{ref7}) and (\ref{ref8}) we have
\begin{eqnarray}\label{ref9}
M_{ef.}^2&:=&V^{ef.}_{,\phi\phi}(0)=V_{,\phi\phi}(0)-\frac{2}{3M_P^4}(\rho_m+\rho_r+\rho_\sigma)(\rho_m\nonumber \\
&+&2\rho_r+\rho_\sigma)f_{,\phi\phi}(0)+
B_{,\phi\phi}(0)\rho_\sigma.
\end{eqnarray}
The above equation is true as long as $\phi$ stays at $\phi=0$, or as long as $M_{ef.}^2 > 0$. Indeed when $M_{ef.}^2 > 0$ the scalar field is trapped at $\phi = 0$, representing a stable point that corresponds to zero dark energy density. As $V^{ef.}_{,\phi}(0)=0$ and $V^{ef.}_{,\phi\phi}(0)>0$, the solution $\phi=-\phi=0$ is the minimum of the effective potential and respects the $Z_2$ symmetry. Note that $M_{ef.}$ might be considered as the mass of small fluctuation around vacuum $\phi=0$ \cite{sym1,sym3}: Substituting  $\phi=0+\delta \phi=\delta\phi$ into the equation of motion and linearizing in $\delta\phi$, we obtain  $\ddot{\delta\phi}+3H\dot{\delta\phi}+V^{ef.}_{,\phi \phi}(0)\delta\phi=0$. A similar proposal has been used in the symmetron model to explain the screening of the scalar field in dense regions \cite{sym3}. To maintain the point $\phi=0$ as a stable point for $a < a_c$, where $M_{ef.}^2(a_c) = 0$, we require $f_{,\phi\phi}(0) < 0$. This condition ensures also insignificance of dark energy for all $a < a_c$. To trigger dark energy evolution, $\phi = 0$ must become an unstable point.
The coefficients of $f_{,\phi\phi}(0)$  are terms like $\rho_i\rho_j$, which dilute faster than the coefficient of $B_{,\phi\phi}(0)$. Thus, setting $B_{,\phi\phi}(0) < 0$  induces generally an instability at $\phi=0$ by the Universe expansion. During the radiation and matter dilution the second derivative of the effective potential becomes negative at $\phi=0$, i.e. $M_{ef.}^2:=V^{ef.}_{,\phi\phi}(0)<0$, indicating that $\phi=0$ is now an unstable point corresponding to the maximum of the potential. So the quintessence becomes tachyonic(see Fig.(\ref{fig.0}) in the following subsection). In this situation a small fluctuation causes the quintessence to roll down its effective potential \cite{sym3}. The resulting non-trivial $Z_2$ violating  solution (i.e. $\phi(t)\neq 0$ and does not respect the $Z_2$ symmetry (see Fig.(\ref{fig.1}))) and its fate is determined by solving the Friedmann and continuity equations. The quintessence evolution depends on its couplings and potential. For instance, a suitable choice might involve a potential that initially rises from zero and gradually flattens, resulting in a slowly rolling quintessence driving the current universe acceleration, while respecting the condition (\ref{17}). We have assumed $V_{,\phi\phi}(0) > 0$ in our model. Without this assumption, the potential turns negative, while a positive acceleration requires a positive potential \cite{fa}.

In the absence of the GB term, this mechanism necessitates $B_{,\phi\phi}(0) > 0$ and $V_{,\phi\phi}(0) < 0$, resulting in a negative potential, which as said is unfavorable for achieving positive acceleration \cite{fa}. This is why within a screening symmetron model, incorporating a cosmological constant becomes necessary to describe the late-time acceleration \cite{sym1, sym2}, unless alternative frameworks like teleparallel gravity are employed \cite{tel}.

\subsection{A numerical solution}

We proceed with a specific example to show how the model works. For the GB coupling we choose a quadratic function,
and for $m$ a Gaussian type function \cite{mass1,mass2}
\begin{equation}\label{21}
f=-\frac{1}{2}\alpha^2\phi^2,\,\,\,m=m_0e^{-\frac{1}{2}\beta^2\phi^2},
\end{equation}
which gives $B=\ln{m_0}-\frac{1}{2}\beta^2\phi^2$. $\alpha$, $\beta$, and $m_0>0$, are real numbers. Note that as the evolution equations only depend on the derivatives of $B$, the mechanism is not directly influenced by the value of $m_0$. As we consider $\sigma$ as a barotropic pressureless fluid, if we  assign it a temperature such as $T$, the inequality  $m_0\ll T$ must hold.
We choose
the quintessence potential as the sum of cosmological constant and a Gaussian type
potential\cite{coup1,coup2,coup3}
\begin{equation}\label{22}
V=V_0(1-e^{-\frac{1}{2}\mu^2\phi^2}).
\end{equation}
 $V_0>0$, and $\mu$ are real numbers and $V_{,\phi\phi}(0)=V_0\mu^2>0$. The minimum is $V(\phi=0)=0$.  For $\mu^2\phi^2\gtrsim 1$ the potential becomes nearly flat and for $\mu^2\phi^2\gg 1$ becomes a cosmological constant $V=V_0$.
 Note that $f$, $m$ and $V$ are even functions as needed for the initial $Z_2$ symmetry.

To perform numerical analysis, we use dimensionless parameters $\hat{H}=\frac{H}{H^*}$, $\hat{t}=H^*t$,  $\hat{\mu}=M_P\mu$, $\hat{\alpha}=H^*\alpha$, $\hat{\beta}=M_P\beta$, $\hat{\rho}=\frac{\rho}{M_P^2{H^*}^2}$, $\hat{V}_0= \frac{V_0}{M_P^2{H^*}^2}$, $\hat{\phi}=\frac{\phi}{M_P}$, where $H^*$ is a scale. So, we may write the Friedmann and continuity equations as follows:
\begin{eqnarray}\label{23}
&&3\hat{H}^2=\frac{1}{2} \hat{\phi}'^2+\hat{V}
_0(1-e^{-\frac{1}{2}\hat{\mu}^2\hat{\phi}^2})-12\hat{\alpha}^2\hat{H}^3\hat{\phi}\hat{\phi}'
+\hat{\rho}_r+\hat{\rho}_d+\hat{\rho}_\sigma\nonumber \\
&&\hat{\phi}''+3\hat{H}\hat{\phi}'+\hat{V}_0\hat{\mu}^2 \hat{\phi}e^{-\frac{1}{2}\hat{\mu}^2\hat{\phi}^2}-12\hat{\alpha}^2(\hat{H}^2+\hat{H}')
\hat{H}^2 \hat{\phi}-\hat{\beta}^2\hat{\phi}\hat{\rho}_\sigma=0,\nonumber \\
&&\hat{\rho}_\sigma'+3\hat{H}\hat{\rho}_\sigma+\hat{\beta}^2\hat{\phi}\hat{\rho}_\sigma\hat{\phi}'=0\nonumber \\
&&\hat{\rho}_r'+4\hat{H}\hat{\rho}_r=0\nonumber\\
&&\hat{\rho}_m'+3\hat{H}\hat{\rho}_m=0,\nonumber\\
\end{eqnarray}
where "prime" denotes the derivative with respect to the dimensionless time $\hat{t}$. In (\ref{23}), $H'$ (including first-order time derivatives) is:
\begin{eqnarray}\label{24}
&&(2+8\hat{\alpha}^2\hat{H}\hat{\phi} \hat{\phi}'+48\hat{\alpha}^4\hat{H}^4\hat{\phi}^2)\hat{H}'=-(1+4\hat{\alpha}^2\hat{H}^2)\hat{\phi}^{'2}+16\hat{\alpha}^2\hat{H}^3
\hat{\phi}\hat{\phi}'\nonumber \\
&&-48\hat{\alpha}^4\hat{H}^6\hat{\phi}^2+4\hat{\alpha}^2\hat{\mu}^2\hat{V}_0\hat{H}^2\hat{\phi}^2
e^{-\frac{1}{2}\hat{\mu}^2\hat{\phi}^2}-4\hat{\alpha}^2\hat{\beta}^2\hat{H}^2\hat{\phi}^2\hat{\rho}_{\sigma}-
\frac{4}{3}\hat{\rho}_r-\hat{\rho}_m \nonumber \\
&&-\hat{\rho}_\sigma.
\end{eqnarray}
We derived (\ref{24}) by combining equations (\ref{7}) and (\ref{9}) as follows.  By substituting $\ddot{f}=f_{,\phi}\ddot{\phi}+f_{,\phi \phi}\dot{\phi}^2$ into the second equation of (\ref{7}), and subsequently utilizing (\ref{9}) to express $\ddot{\phi}$ in terms of first-order time derivatives, we arrived at (\ref{24}).

\begin{figure}[H]
\centering
\includegraphics[height=6cm]{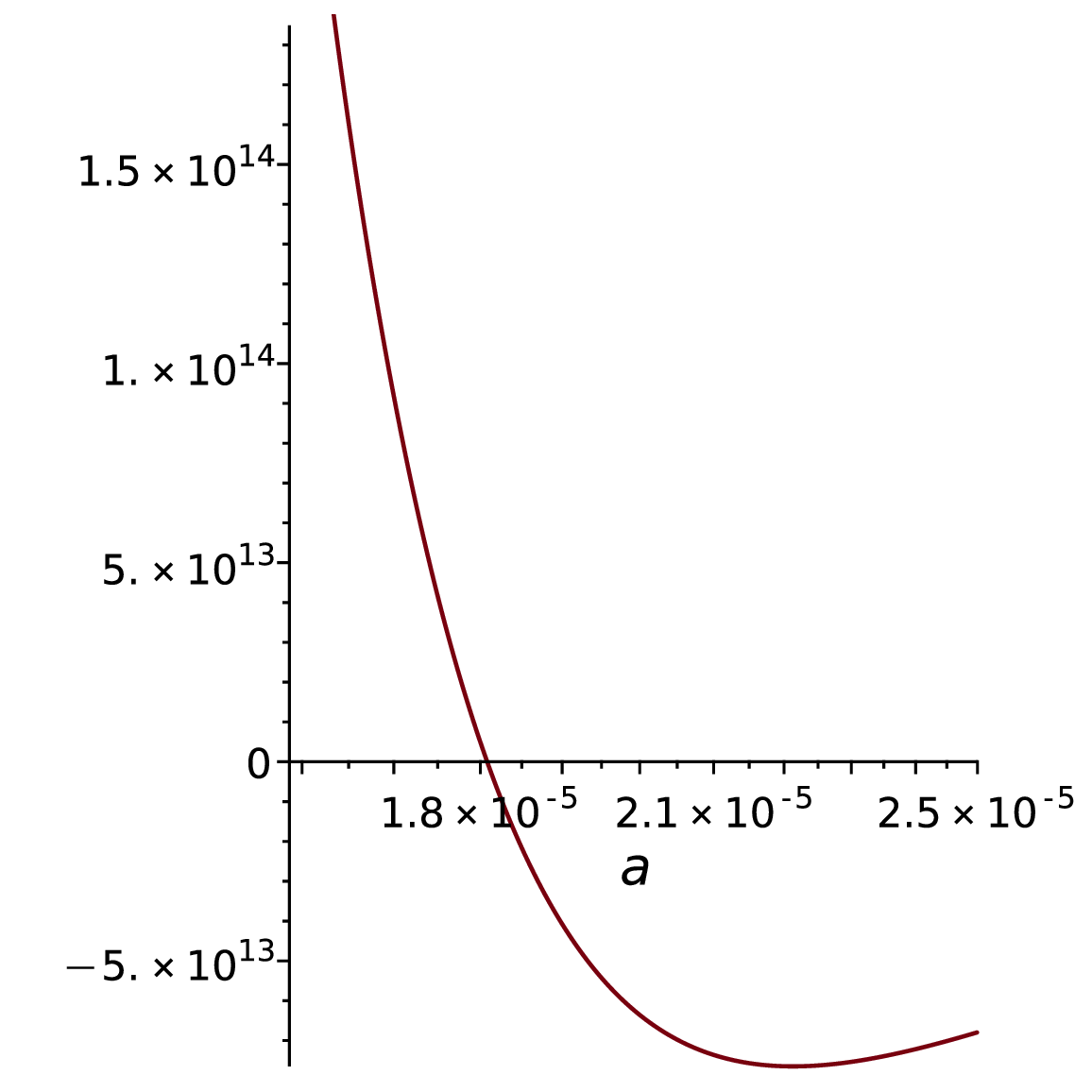}
\caption{Dimensionless effective mass squared, ${\hat{M}_{ef.}}^2$, in terms of the scale factor }
\label{fig.0}
\end{figure}

Following our discussion after (\ref{ref9}), $\phi=0$ becomes an unstable point when densities satisfy:
\begin{equation}\label{25}
\hat{\mu}^2 \hat{V}_0-\hat{\beta}^2\hat{\rho}_{\sigma}+\frac{2}{3}(\hat{\rho}_m+
\hat{\rho}_\sigma+\hat{\rho}_r)(\hat{\rho}_m+\hat{\rho}_\sigma+2\hat{\rho}_r)\hat{\alpha}^2<0,
\end{equation}
Before this, neither the scalar field nor the GB term entered the Friedmann's equations, and the dark energy density was zero.
We take the initial point at $a=\frac{1}{55000}$ (redshift $z\simeq 55000$) in the radiation dominated and set the initial conditions as
 \begin{eqnarray}\label{26}
\hat{\phi}(0)=0,  \hat{\phi}'(0)=10^{-20}, \Omega_\rho=\frac{250}{344}\simeq 4.9\%, \nonumber\\
\Omega_\sigma=\frac{17}{344}\simeq  22.4\%,  \Omega_r=\frac{77}{344}\simeq 72.7\%.
\end{eqnarray}
The relative density is defined by $\Omega_i=\frac{\hat{\rho}_i}{3\hat{H}^2}=\frac{\rho_i}{3M_P^2H^2}$.
We choose the parameters as:
\begin{equation}\label{27}
\hat{\mu}=5.5, \hat{\beta}=2.6, \hat{V}_0=2.052, \hat{\alpha}=2.028\times 10^{-8},  H^{*}=H_0,
\end{equation}
where $H_0=67,4\pm 0.5$  kms$^{-1}$Mpc$^{-1}$ is the present Hubble parameter \cite{Planck}. In (\ref{27}), we have chosen the parameters such that $\hat{V}_0$ (as we will discuss) is compatible with the relative cosmological constant dark energy density in $\Lambda$CDM model. With the very small value of GB-quintessence coupling $\alpha$ and the values of order of magnitude one that we have chosen for $\beta$ and $\mu$, the square of the effective mass gains a root ($a_c$) in the radiation dominated era such that $\frac{dM_{ef.}^2}{da}(a_c)<0$, leading to $V^{ef.}_{,\phi\phi}(0)<0$ for $a>a_c$, in accordance with our discussion after (\ref{ref9}). Note that our choice for $\alpha$ and the initial conditions (\ref{26}) is also consistent with (\ref{25}). The dimensionless effective mass squared, ${\hat{M}_{ef.}}^2:=\frac{M_{ef.}^2}{H_0^2}$ is depicted in Fig.(\ref{fig.0}).
$\phi=0$ is a stable point as long as $M_{ef.}^2=V^{ef.}_{,\phi\phi}(0)>0$, but for $a>a_c$, becomes an unstable point $V^{ef.}_{,\phi\phi}(0)<0$.

\begin{figure}[H]
\centering
\includegraphics[height=6cm]{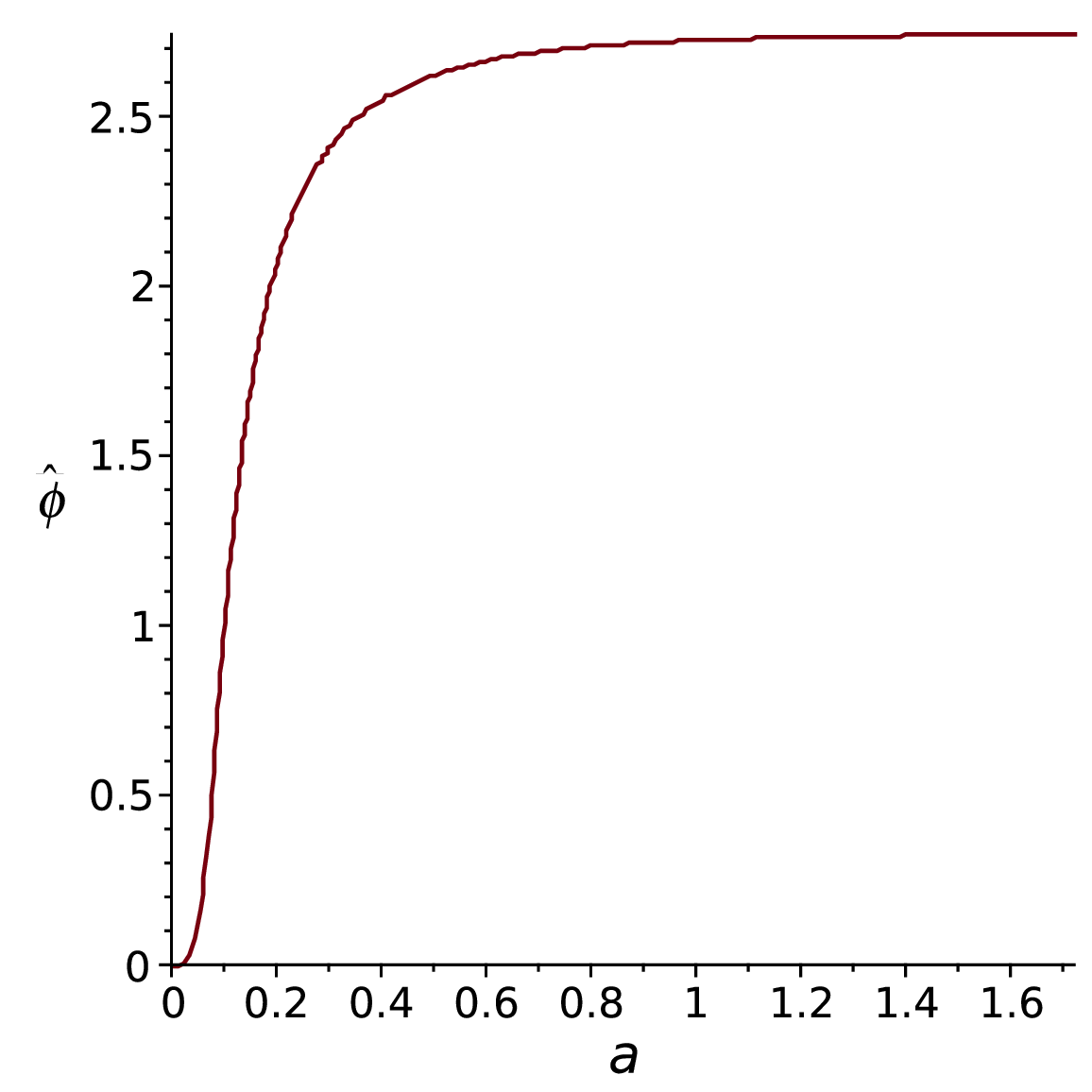}
\caption{Quintessence, $\hat{\phi}$, evolution in terms of the scale factor}
\label{fig.1}
\end{figure}

\begin{figure}[H]
\centering
\includegraphics[height=6cm]{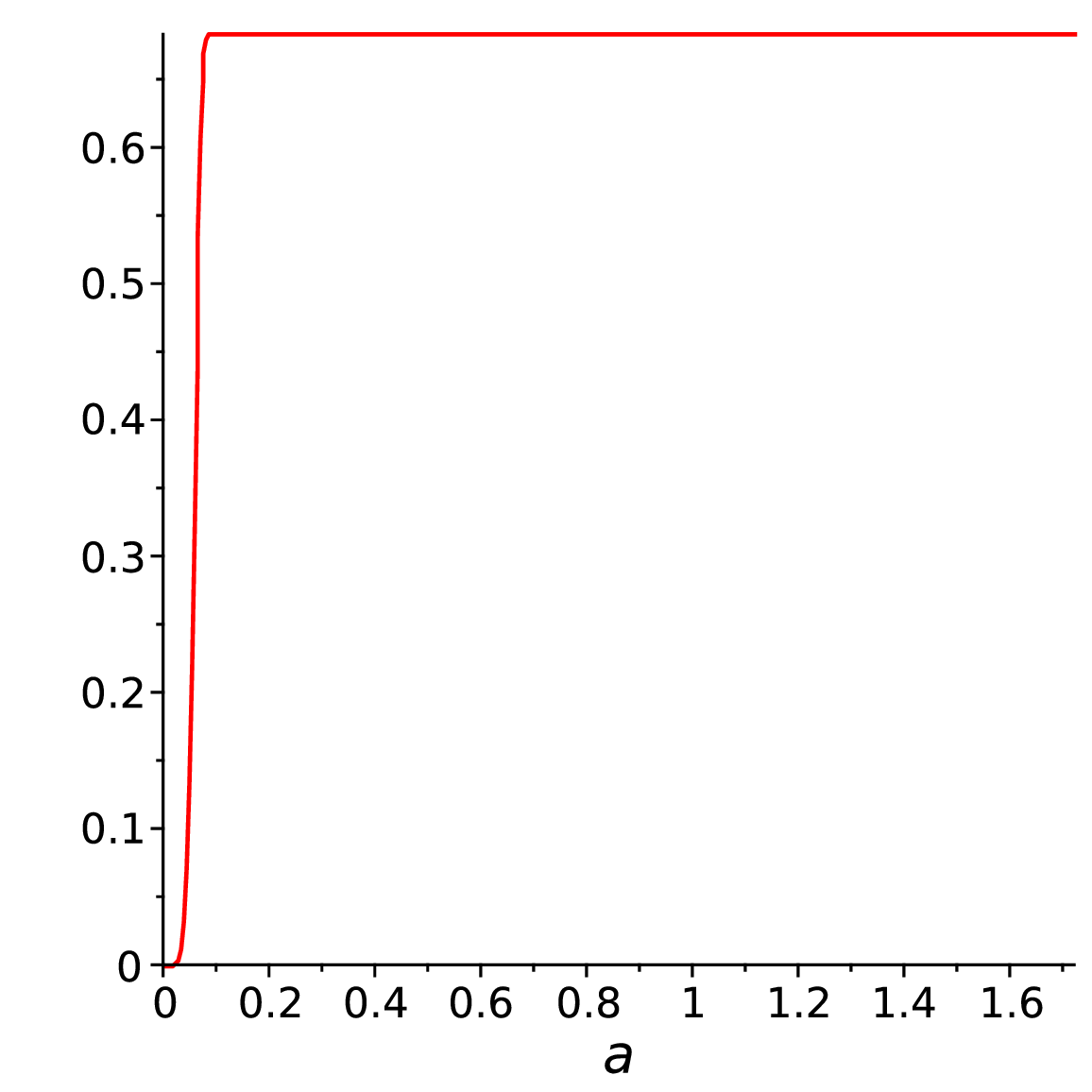}
\caption{$\frac{1}{3}\hat{V}_0 (1-e^{-\frac{1}{2}\mu^2\phi^2})$ in terms of the scale factor}
\label{fig.2}
\end{figure}

In Fig.(\ref{fig.1}), using the system of equations (\ref{23}), we have depicted the evolution of the quintessence.
The quintessence begins its evolution from  $\phi=0$, when $M_{ef.}^2<0$. Despite the activation of the scalar field during the period of radiation dominance, due to friction, it actually increases significantly after $a\sim 0.01$ in the matter-dominated era and eventually slowly rolls along the flat potential depicted in Fig.(\ref{fig.2}). the quintessence's slow rolling also helps to satisfy (\ref{17}) and also retrieving a similar model to $\Lambda$CDM.
 Fig.(\ref{fig.2}) shows that the potential (\ref{22}) is actually zero until the matter-dominated era, and then increases and becomes flat when $\mu^2 \phi^2\gg 1$, behaving eventually like a cosmological constant $\frac{1}{3}\hat{V}=\frac{V}{3M_P^2H_0^2}=0.68$, which is compatible with the relative dark energy density : $\Omega_\Lambda=0.6889\pm 0.0056$
for $\Lambda$CDM model, reported based on Planck TT,TE,EE+lowE+lensing (68\% CL) in \cite{Planck}. Note that at this stage, the GB term has no role in the evolution of the Universe.

\begin{figure}[H]
\centering
\includegraphics[height=6cm]{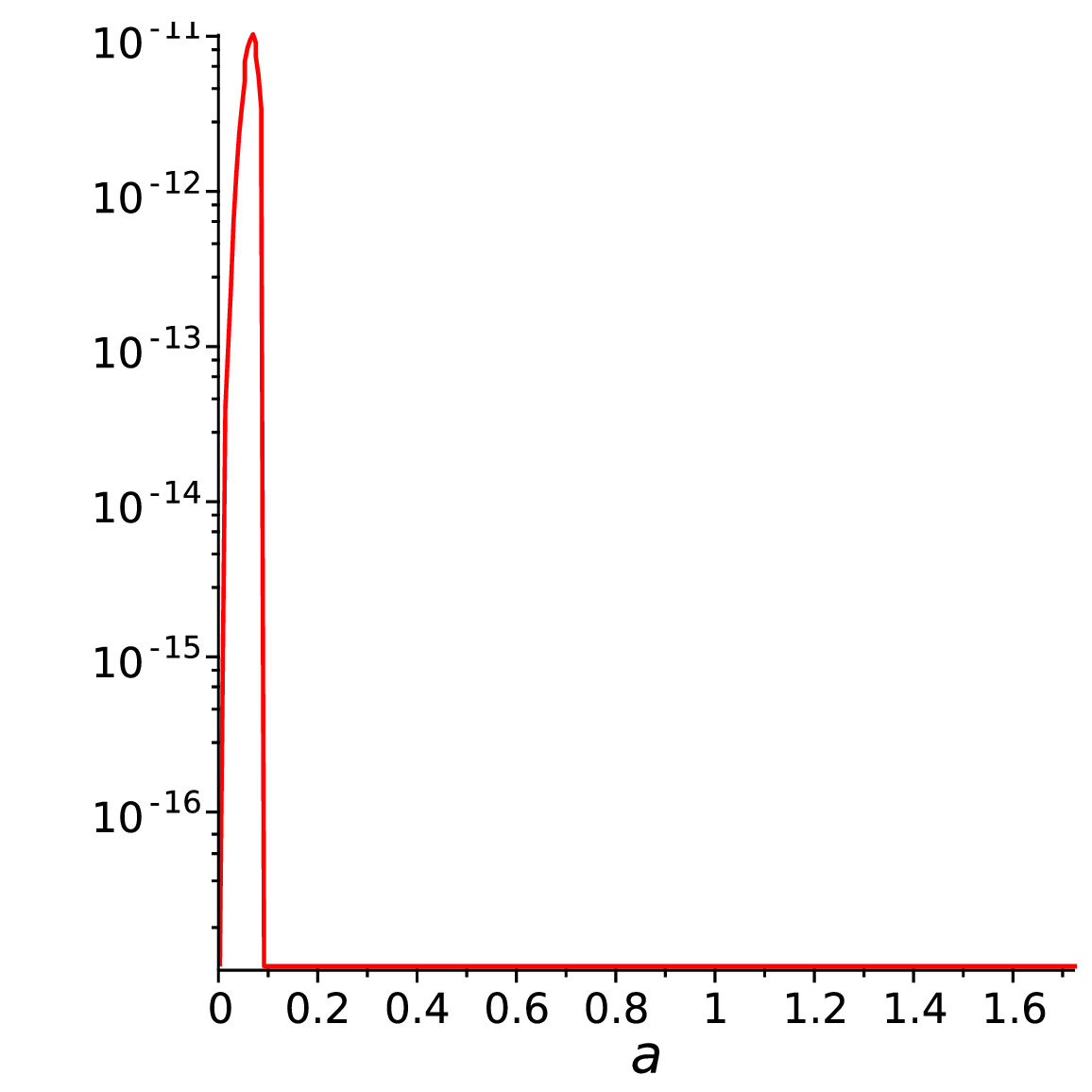}
\caption{$1-c_T$ in terms of the scale factor }
\label{fig.3}
\end{figure}

Now, let us investigate the assumed condition (17). $1-c_T$, where
\begin{equation}\label{28}
c_T^2=\frac{4{\hat{\alpha}}^2{\hat{\phi}}'^2+4{\hat{\alpha}}^2\hat{\phi}{\hat{\phi}}''+1}{4\hat{\alpha}^2{\hat{\phi}}'\hat{\phi}+1},
\end{equation}
is depicted in Fig.(\ref{fig.3})
Near $a=1$, the condition is satisfied. This is thanks mainly to the tiny value selected for $\hat{\alpha}$ and also to the slow rolling of the quintessence in the late time. In the period $a<a_c$, when the quintessence remained at the stable point, $c_T$ was the same as the light speed $c$, when the quintessence becomes active $c_T$ deviates from $c$, and eventually as the potential becomes nearly flat, the quintessence becomes again a constant and the $c_T$ equals the light speed. By the Rolle's theorem, there is a point for which $dc_T/da=0$, corresponding to the peak in Fig.(\ref{fig.3}). The faster the field reaches the final constant value, the sharper this peak will be.

\begin{figure}[H]
\centering
\includegraphics[height=6cm]{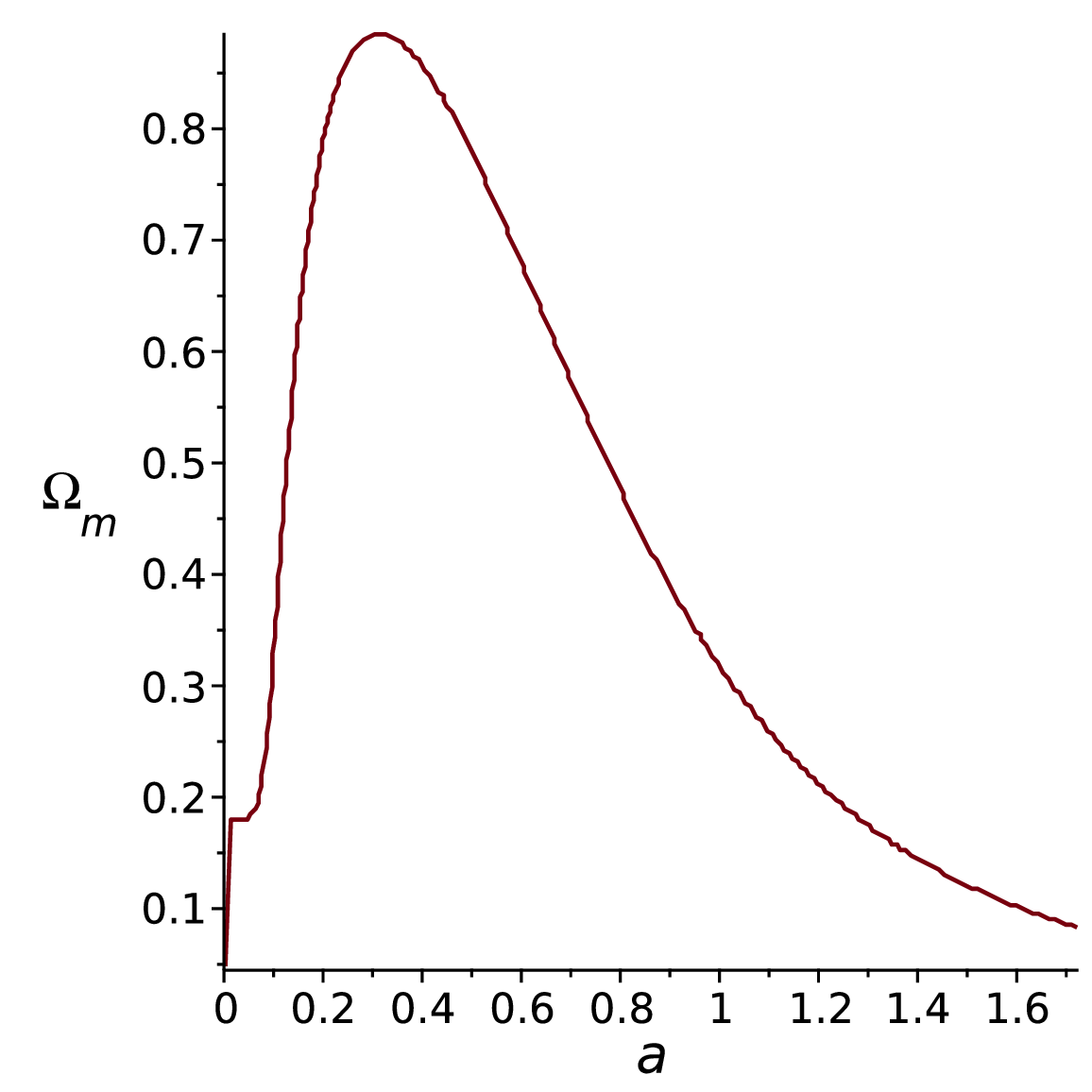}
\caption{$\Omega_m$ in terms of the scale factor }
\label{fig.4}
\end{figure}

\begin{figure}[H]
\centering
\includegraphics[height=6cm]{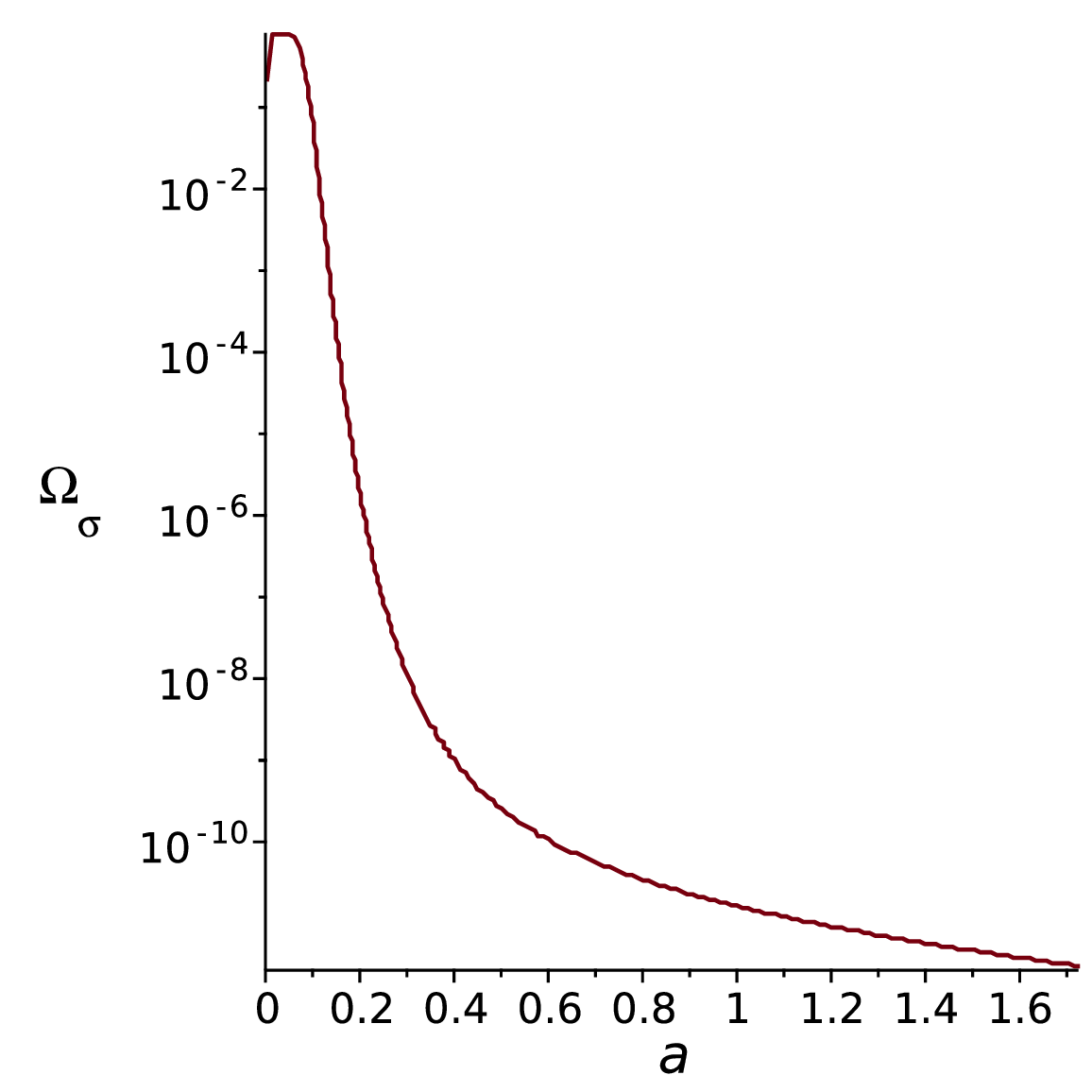}
\caption{$\Omega_\sigma$ in terms of the scale factor }
\label{fig.5}
\end{figure}

In Fig.(\ref{fig.4}), we have depicted the relative density $\Omega_m$.
At the present time, $a=1$,  we obtain $\Omega_m=0.32$, which is compatible with the relative matter density reported in \cite{Planck}, i.e.  $\Omega_m=0.3111\pm 0.0056$ (based on  Planck TT,TE,EE+lowE+lensing at 68\% CL). The decrease in non-interacting matter density occurs as a result of redshift, whereas $\sigma$ dark matter directly interacts with quintessence. Consequently, we anticipate distinct behavior from it. The depiction of $\Omega_\sigma$ can be found in Fig.(\ref{fig.5}).
 This indicates that the interaction between $\sigma$ dark matter and the quintessence has significantly decreased the density, several times more than the redshift did, albeit at the expense of increasing the energy density of the scalar field. At the present time we obtain $\Omega_\sigma(a=1)=2.1\times 10^{-11}$. If all dark matter interacted with the field in this way, its density would be significantly reduced, which is inconsistent with observations. This was our reason for separating the interacting sector.

\begin{figure}[H]
\centering
\includegraphics[height=6cm]{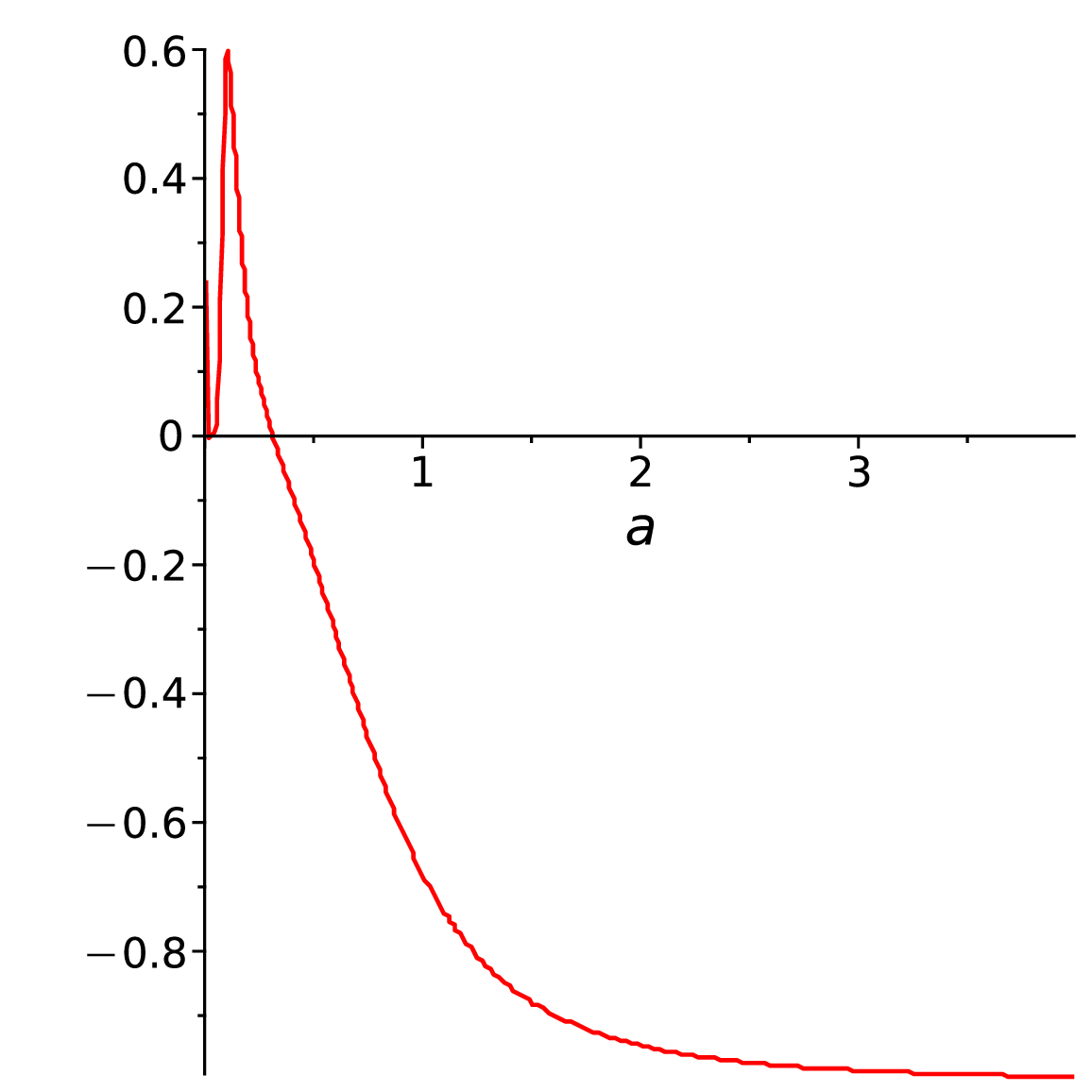}
\caption{EoS parameter of the Universe in terms of the scale factor }
\label{fig.6}
\end{figure}

Finally, the equation of state (EoS) of the Universe,
\begin{equation}\label{eos}
w=-1-\frac{2}{3}\frac{\hat{H}'}{\hat{H}^2}=-1-\frac{2}{3}\frac{\dot{H}}{H^2},
\end{equation}
is depicted in fig.(\ref{fig.6}),
In this example the Universe entered the acceleration phase at $a=0.66$ or $z=0.52$. In this figure, at the present epoch $a=1$, the EoS of the Universe is $w=-0.69$.
Using the Friedmann equations (\ref{ref4}), for the late time when the Universe is dominated by dark energy, baryonic matter and cold dark matter, we can rewrite (\ref{eos}) as
\begin{eqnarray}
w&=&-1+\frac{\rho_m+(1+w_d)\rho_d}{\rho_m+\rho_d}\nonumber \\
&=&w_d\Omega_d,
\end{eqnarray}
where $w_d$ is the EoS of dark energy, and $\Omega_d$ is its relative density. Since $w=-0.69$, and $\Omega_d=0.68$, we deduce that $w_d= -1.01$, which
is compatible with the result reported in \cite{Planck} for dark energy EoS parameter $w_0=-1.03\pm 0.03$. For $z<0$, the EoS of the Universe tends to $w=-1$.

 The qualitative behavior observed in figures such as Fig.(\ref{fig.4}) directly arises from the underlying structure of our proposed model. This structure encompasses the form of the potential and coupling, along with the sign of their parameters, as explained before this subsection. The choice of parameter magnitudes within the model was made to adhere to the imposed  conditions (\ref{17}), leading to activation of dark energy in the matter dominated period. But,to be more precise in our numerical example, adjustment of the parameters have been carried out to make the results compatible with the observational data reported in \cite{Planck}. Note that as our equations in the numerical analysis are in terms of dimensionless parameters, all sets $\{\alpha, \rho_{initial}, V_0, H^*\}$ resulting a same $\{\hat{\alpha},\hat{\rho}_{initial},\hat{V}_0\}$ give the same results.

\section{Conclusion}

The Scalar-Gauss-Bonnet (SGB) model has been one of the popular models in studying the recent acceleration of the Universe.  In this study, we introduced a novel role for the GB term, i.e. maintaining the quintessence at a stable fixed point before the matter-dominated era corresponding to a state with zero dark energy density. This gives a possible explanation for the insignificance of the relative early dynamical dark energy density, as pointed out in many papers and required for structure formation. Also by postponing the rising of dark energy until the matter-dominated era, the model explains why despite the slower dilution rate of dark energy with respect to matter, they can have similar magnitudes today.

To model the above mechanism, we considered a quintessence coupled to the GB invariant and conformally coupled to a dark matter sector, where the latter can be interpreted in the context of mass-varying dark matter (see (\ref{9}) and (\ref{dm}) and the corresponding discussions). The conformal coupling has been previously used in screening scalar-tensor models to explain the absence of dark energy in dense regions by restoring a $Z_2$ symmetry \cite{sym3}. In the SGB model with a conformally coupled dark matter, the Friedmann and quintessence equations of motion are modified (see Eqs. (\ref{7}) and (\ref{9})). These modifications can be used to define an effective potential for the quintessence, which depends on matter and radiation densities (see (\ref{ref7})). When the scale factor passes the critical value $a_c$, the initial stable point (minimum point of the effective potential) becomes an unstable point ( maximum point of the effective potential (see (\ref{ref9}) and its following discussion)). Hence, by a small fluctuation, the quintessence rolls down its effective potential and its evolution begins\cite{sym1,sym2}. The fate of this evolution depends on the coupling and potential. In contrast to \cite{sym1,sym2}, the presence of the GB coupling allows to obtain a positive potential for the quintessence  which is required for the late time acceleration (see the discussion after (\ref{ref9})).

We have demonstrated that, under the assumption that the speed of gravitational wave at low redshifts is very close to the speed of light, it is advantageous to consider $a_c$ during the radiation-dominated era. This choice allows for the selection of a tiny scalar-GB coupling coefficient, as discussed after equation (\ref{27}). Moreover, this assumption supports the presence of a quintessence field that evolves very slowly at low redshifts, as indicated in equation (\ref{16}). In our numerical example, we opted for a potential that eventually becomes flat, as illustrated in Fig. (\ref{fig.2}). Consequently, the model adopts a behavior akin to the $\Lambda$CDM model in the late time.

\end{document}